 \documentclass[preprint ]{aastex} 

\shorttitle{Fe~XVII $I(17.10~{\rm \AA })/I(17.05~{\rm \AA })$ Density
Diagnostic}
\shortauthors{Mauche, Liedahl, \& Fournier}

 \slugcomment{Accepted for publication in 
             {\it Ap.J.\/} June 27, 2001}

\newcommand\Mdot  {\dot{M}}
\newcommand\Mwd   {M_{\rm wd}}
\newcommand\Rwd   {R_{\rm wd}}
\newcommand\Msun {{\rm M_{\odot}}}
\newcommand\lax{{\lower0.75ex\hbox{ $<$ }\atop\raise0.5ex\hbox{ $\sim$ }}}
\newcommand\gax{{\lower0.75ex\hbox{ $>$ }\atop\raise0.5ex\hbox{ $\sim$ }}}

\begin{document}

\title{First Application of the \ion{Fe}{17} 
I(17.10~{\rm \AA })/I(17.05~{\rm \AA }) Line Ratio \\
to Constrain the Plasma Density of a Cosmic X-ray Source}

\author{Christopher W.\ Mauche, Duane A.\ Liedahl,
    and Kevin B.\ Fournier}
\affil{Lawrence Livermore National Laboratory,
       L-43, 7000 East Avenue, Livermore, CA 94550; \\
       mauche@cygnus.llnl.gov, duane@virgo.llnl.gov, fournier2@llnl.gov}



\begin{abstract}

We show that the \ion{Fe}{17} $I(17.10~{\rm \AA })/I(17.05~{\rm \AA })$
line ratio observed in the {\it Chandra\/} HETG spectrum of the
intermediate polar EX~Hydrae is significantly smaller than that observed
in the Sun or other late-type stars. Using the Livermore X-ray Spectral
Synthesizer, which calculates spectral models of highly charged ions
based on HULLAC atomic data, we find that the observed $I(17.10~{\rm \AA
})/I(17.05~{\rm \AA })$ line ratio can be explained if the plasma density
$n_{\rm e}\gax 3\times 10^{14}~\rm cm^{-3}$. However, if photoexcitation
is included in the level population kinetics, the line ratio can be
explained for any density if the photoexcitation temperature $T_{\rm bb}
\gax 55$~kK. For photoexcitation to dominate the \ion{Fe}{17} level
population kinetics, the relative size of the hotspot on the white dwarf
surface must be $f\lax 2$\%. This constraint and the observed X-ray flux
requires a density $n\gax 2\times 10^{14}~\rm cm^{-3}$ for the post-shock
flow. Either way, then, the {\it Chandra\/} HETG spectrum of EX~Hya
requires a plasma density which is orders of magnitude greater than that
observed in the Sun or other late-type stars.

\end{abstract}

\keywords{atomic processes ---
          binaries: close ---
          stars: individual (EX Hydrae) ---
          stars: magnetic fields ---
          Sun: corona ---
          X-rays: binaries}

 

\section{Introduction}

Because of the high abundance of Fe, the persistence of Ne-like Fe over
a broad temperature range ($T_{\rm e}\approx 2$--12 MK), and the large
collision strengths for $2p\to nd$ transitions, the $2p^6$--$2p^53l$
($l=s$, $d$) lines of \ion{Fe}{17} at 15--17~\AA \ are prominent in the
X-ray spectra of high-temperature plasmas found in settings as diverse
as tokamaks \citep{kla78, phi97, bei01}, the Sun \citep{rug85, phi97,
sab99}, and both late- and early-type stars \citep{can00, ayr01, kah01,
wal01}. The importance of \ion{Fe}{17} has engendered numerous studies
of its atomic structure and level population kinetics, usually with
an emphasis on the {\it temperature\/} dependence of line ratios within
the $2p$--$3s$ and $2p$--$3d$ multiplets \citep{rug85, smi85, ray86}.
In this paper, we focus on the {\it density\/} dependence of the
\ion{Fe}{17} X-ray spectrum.

The lowest-lying configurations of \ion{Fe}{17} are the $2p^6$ ground
state, 
the $2p^53s$ multiplet (4  levels, $E=725.3$--739.0 eV),
the $2p^53p$ multiplet (10 levels, $E=754.3$--789.6 eV), and
the $2p^53d$ multiplet (12 levels, $E=800.3$--826.0 eV).
Because electron impact excitation from the ground state is strong for 
transitions into the $2p^5nd$ configurations, but weak for transitions
into the $2p^53s$ configuration, the population flux into the 
lowest-lying configuration is dominated by radiative cascades originating
on higher-lying energy levels, rather than direct excitation from the
ground state \citep{lou73}. The four levels of the $2p^53s$ configuration
have, in increasing energy order, total angular momenta $J=2$, 1, 0, and
1. The $2p^6$--$2p^5 3s\> (J=0)$ transition is strictly forbidden, but
the remaining three levels $J=2$, 1, and 1 decay to ground, producing
lines at 17.10, 17.05, and 16.78~\AA , respectively. The 17.10~\AA \ line
is produced by a magnetic quadrupole transition, but it is nevertheless
bright because the upper level is populated efficiently by radiative
cascades, and its radiative branching ratio to ground is 1.0. Since
the radiative decay rate of this transition is slow compared to the
other $2p$--$3l$ lines, collisional depopulation sets in at lower
densities. Thus, the intensity ratio of the 17.10~\AA\ line to any of
the other $2p$--$3l$ lines provides a diagnostic of the plasma density,
as was first pointed out by \citet{kla78}.

Of these density-dependent line ratios, we study primarily the
$I(17.10~{\rm \AA })/I(17.05~{\rm \AA })$ line ratio, which has a
critical density of a ${\rm few} \times 10^{13}~\rm cm^{-3}$. To utilize
this diagnostic in a cosmic X-ray source, we require an instrument with
sufficient spectral resolution to cleanly resolve the two lines, and a
bright X-ray source with high characteristic densities, intrinsically
narrow lines, and low interstellar absorption. The instrumental
requirements are supplied by the High-Energy Transmission Grating
Spectrometer \citep[HETGS;][]{mar94} aboard the {\it Chandra\/} X-ray
Observatory, and the source requirements are satisfied by nearby magnetic
cataclysmic variables (CVs): semidetached binaries composed of a magnetic
($B\sim 1$--100~MG) white dwarf primary and a low-mass secondary. The
global properties are different for polars (AM Her stars) and
intermediate polars (DQ~Her stars), but in both classes of magnetic
CVs the flow of material lost by the secondary is channeled onto small
spots on the white dwarf surface in the vicinity of the magnetic poles.
Before reaching the stellar surface, the supersonic [$v_{\rm ff}
=(2G\Mwd/\Rwd)^{1/2} \approx 3600~\rm km~s^{-1}$] flow passes through a
strong shock [$T_{\rm s} =3G\Mwd \mu m_{\rm H}/ 8\Rwd \approx 200$~MK],
where most of its kinetic energy is converted into thermal energy. For a
mass-accretion rate $\Mdot =10^{15}~\rm g~s^{-1}$ ($L_{\rm X}=G\Mwd
\Mdot/2\Rwd\approx 3\times 10^{31}~\rm erg~s^{-1}$) and a relative spot
size $f=0.1$, the density of the flow immediately behind the shock is
$n=\Mdot/4\pi f\Rwd ^2\mu m_{\rm H} (v_{\rm ff}/4)\approx 10^{13}~\rm
cm^{-3}$. As the flow settles onto the white dwarf, it cools and its
density increases, so the density is significantly higher where the
temperature reaches a few million degrees, where \ion{Fe}{17} is the
dominant charge state of Fe. Doppler broadening of the emission lines
is reduced because of the ordered, quasi-radial geometry, and the
ever-decreasing velocity ($v < v_{\rm ff} /4\approx 900~\rm km~s^{-1}$,
hence $\Delta\lambda/\lambda < 3\times 10^{-3}$) of the post-shock flow. 

With its bright X-ray emission lines and low interstellar absorption,
the well-studied \citep{hel87, ros88, hur97, fuj97, all98, mau99}
intermediate polar EX~Hydrae is an ideal magnetic CV in which to study
high-density plasmas. Spectroscopic evidence for high densities in
this source is provided by \citet{hur97}, who used the ratio of the
\ion{Fe}{20}/\ion{Fe}{23} 133~\AA \ blend to the \ion{Fe}{21} 129~\AA \
line observed in an {\it EUVE\/} SW spectrum to infer a density $n_{\rm
e}\gax 10^{13}~\rm cm^{-3}$ for the $T_{\rm e}\sim 10$~MK plasma in
EX~Hya. Below, we use the \ion{Fe}{17} $I(17.10~{\rm \AA })/I(17.05~{\rm
\AA })$ line ratio observed in a {\it Chandra\/} HETG spectrum to infer a
density $n_{\rm e}\gax 3\times 10^{14}~\rm cm^{-3}$ for the $T_{\rm e}
\sim 4$~MK plasma in EX~Hya. We discuss the observations and analysis in
\S 2, \ion{Fe}{17} spectral models in \S 3, and close with a summary and
discussion in \S 4.

\section{Observations and Analysis}

The {\it Chandra\/} HETG/ACIS-S observation of EX~Hya was performed
between 2000 May 18 $\rm 9^h41^m$ and May 19 $\rm 2^h54^m$ UT for a total
exposure of 59.1 ks. Extraction of the grating spectra and calculation
of the effective area files was accomplished with the CIAO 2.1 suite of
software using the reprocessed data products and new calibration data
files (version R4CU5UPD8) for sequence 300041. A preliminary discussion of
the resulting Medium Energy Grating (MEG) and High Energy Grating spectra
has been presented by \citet{mau00a} and \citet{mau00b}; here we discuss
only the MEG spectrum as it bears on the \ion{Fe}{17} emission spectrum.

A detail of the MEG spectrum from 14.5 to 17.5~\AA \ in shown in Figure~1.
This representation of the spectrum, in $\rm counts~s^{-1}~\AA ^{-1}$,
combines $\pm 1$st orders and is binned by a factor of two to $\Delta
\lambda =0.01$~\AA . As indicated along the top of the figure, the
spectrum contains emission lines of \ion{Fe}{17}, \ion{Fe}{18}, and
\ion{O}{8}. The striking aspect of this spectrum is the absence of the
\ion{Fe}{17} 17.10~\AA \ emission line, a feature seen in X-ray spectra
of the Sun \citep{rug85, phi97, sab99} and other late-type stars
\citep{can00, ayr01}. 

To determine the flux in the \ion{Fe}{17} emission lines, we fit the MEG
spectrum over the 14.72--15.55~\AA \ plus 16.68--17.20~\AA \ wavelength
interval with a model consisting of a linear background (to account for
the thermal bremsstrahlung continuum obvious in the broadband spectrum)
and eight Gaussians (to account for the six \ion{Fe}{17} lines and two
intervening \ion{O}{8} Lyman lines). Assuming a common Gaussian width
and offset, the fitting function
$f(\lambda ;\vec a)= a_1+ a_2(\lambda-\lambda_0) +\sum _{i=5}^{12}
a_i\exp[-(\lambda-\lambda _i-a_3)^2 /2a_4^2]$.
Fitting the MEG spectrum in $\Delta\lambda =0.005$~\AA \ bins, separately
accounting for $\pm 1$st orders, results in 540 data points and 528
degrees of freedom. For the wavelengths $\lambda _i$, we assume the
values measured by \citet{bro98} for \ion{Fe}{17} and tabulated by
\citet{kel87} for \ion{O}{8}. The resulting fit, combining $\pm 1$st
orders, is shown by the thick gray curve in Figure~1. It gives $\rm
\chi^2/dof=358/528=0.7$ and fit parameters as follows: wavelength offset
$a_3=4.4\pm 0.1$ m\AA , Gaussian width $a_4=0.74\pm 0.14$ m\AA , and
line fluxes 
$I(15.01~{\rm \AA })=3.1\, (\pm 0.2)\times 10^{-4}$,
$I(15.26~{\rm \AA })=1.3\, (\pm 0.2)\times 10^{-4}$,
$I(15.45~{\rm \AA })=1.3\, (\pm 1.5)\times 10^{-5}$,
$I(16.78~{\rm \AA })=2.5\, (\pm 0.2)\times 10^{-4}$,
$I(17.05~{\rm \AA })=4.2\, (\pm 0.3)\times 10^{-4}$, and
$I(17.10~{\rm \AA })=2.0\, (\pm 1.6)\times 10^{-5}~\rm
photons~cm^{-2}~s^{-1}$. Both the wavelength offsets and line widths are
consistent with zero, given the absolute wavelength accuracy ($\pm 11$
m\AA ) and spectral resolution ($\Delta\lambda = 23$~m\AA ) of the MEG.

With six line fluxes we can form five independent line ratios, and to
compare our values to those measured in the Sun we follow \citet{phi97}
and \citet{sab99} and use the 16.78~\AA \ line as the reference; other
line ratios of interest are $I(15.26~{\rm \AA })/I(15.01~{\rm \AA })$ and 
$I(17.10~{\rm \AA })/I(17.05~{\rm \AA })$. These photon flux line ratios
are listed in the middle column of Table~1. It is important to note that
these ratios are virtually unaffected by interstellar absorption: for
$N_{\rm H}= 10^{20}~\rm cm^{-2}$, the {\it largest\/} correction amounts
to only 1\%. For comparison, the \ion{Fe}{17} photon flux line ratios
for 33 solar active regions measured by \citeauthor{sab99} with the Flat
Crystal Spectrometer aboard the {\it Solar Maximum Mission\/} are listed
in the right column of Table~1 (weighted mean and standard deviation).
The $R(\lambda)\equiv I(\lambda)/I(16.78~{\rm \AA })$ photon flux ratios
for EX~Hya and the Sun are plotted in the middle panel of Figure~3. It is
seen that the line ratios measured in EX~Hya and the Sun are consistent
with each other, with the exception of the $R(17.10~{\rm \AA })$ line
ratio [or, equivalently, the $I(17.10~{\rm \AA })/ I(17.05~{\rm \AA })$
line ratio], which is far smaller in EX~Hya.

\section{Model Spectra}

From the results of \citet{kla78}, the small $I(17.10~{\rm \AA})/
I(17.05~{\rm \AA})$ line ratio observed in EX~Hya implies a plasma
density in excess of a ${\rm few}\times 10^{13}~\rm cm^{-3}$. To
quantify this conclusion, we constructed model spectra for \ion{Fe}{17}
using the Livermore X-ray Spectral Synthesizer (LXSS), a suite of IDL
codes which calculates spectral models of highly charged ions based on
Hebrew University/Lawrence Livermore Atomic Code (HULLAC) atomic data.
HULLAC calculates atomic wavefunctions, level energies, radiative
transition rates, and oscillator strengths according to the fully
relativistic, multiconfiguration, parametric potential method
\citep{kla71, kla77}. Electron impact excitation rate coefficients are
computed quasi-relativistically in the distorted wave approximation
\citep{bar88} assuming a Maxwellian velocity distribution. Our
\ion{Fe}{17} model includes radiative transition rates for E1, E2, M1,
and M2 decays and electron impact excitation rate coefficients for levels
with principal quantum number $n\le 6$ and azimuthal quantum number $l\le
4$ for a total of 281 levels. Using these data, LXSS calculates the level
populations for a given temperature and density assuming
collisional-radiative equilibrium. The line intensities are then simply
the product of the level populations and the radiative transition rates.

In addition to collisional excitation, photoexcitation must be
accounted for in the calculation of atomic level populations in UV-bright
sources like early-type stars and compact binaries. For example, in
He-like ions, photoexcitation competes with collisional excitation to
depopulate the  $1s2s\> ^3S_1$ level, leading to the conversion of the
$1s^2\> ^1S_0$--$1s2s\> ^3S_1$ forbidden line $f$ into the
$1s^2\> ^1S_0$--$1s2p\> ^3P_{2,1}$ intercombination blend $i$
\citep{gab69, blu72}. Thus, if the radiation field is strong enough
at the appropriate wavelengths, the $f/i$ line ratio can be in the
``high-density limit'' ($f/i\approx 0$) regardless of the density.
Photoexcitation has been shown to explain the low $f/i$ line ratios of
early-type stars \citep{kah01, wal01}, and could explain the low $f/i$
and \ion{Fe}{17} $I(17.10~{\rm \AA})/I(17.05~{\rm \AA})$ line ratios
observed in EX~Hya.

To account for photoexcitation, we included in the LXSS population
kinetics calculation the photoexcitation rates $(\pi e^2/m_ec) f_{ij}
F_\nu (T)$, where $F_\nu (T)$ is the continuum spectral energy
distribution and $f_{ij}$ are the oscillator strengths of the various
transitions. For simplicity, we assume that $F_\nu (T) = (4\pi /h\nu )
B_\nu (T_{\rm bb})$ (i.e., the radiation field is that of a blackbody
of temperature $T_{\rm bb}$) and the dilution factor of the radiation
field is equal to one-half (i.e., the X-ray emitting plasma is in close
proximity to the source of the photoexcitation continuum). Both of these
assumptions tend to overestimate the importance of photoexcitation.
First, the actual intrinsic spectrum of EX~Hya probably has a strong
break at the Lyman limit, whereas a blackbody does not. This difference
affects transitions with energy spacings $\Delta E > 13.6$ eV such as the 
$1s2s\> ^3S_1$--$1s2p\> ^3P_{2,1}$ ($f\to i$) transitions in Si and S,
but not those of O, Ne, and Mg. Second, the UV--EUV continuum in EX~Hya
must be due to both the accretion disk and the accretion-heated surface
of the white dwarf. By assuming that the dilution factor is equal to
one-half, we are in effect assuming that the accretion-heated surface
of the white dwarf is the dominant source of the UV--EUV light. This
assumption is justified on the basis of the similarity of the shape and
intensity of the FUV spectra of EX~Hya (with a disk) and AM~Her (without
a disk) \citep{mau99}. With these assumptions, we find for a blackbody
temperature $T_{\rm bb}= 30$~kK (see below) that all of the He-like
$f/i$ line ratios through \ion{Si}{13} are significantly affected by
photoexcitation, whereas the \ion{Fe}{17} $I(17.10~{\rm \AA})/
I(17.05~{\rm \AA})$ line ratio is not.

To understand this result, we show in Figure~2 a schematic of the
important level population processes for the $2p^53s\> (J=2)$ level
of \ion{Fe}{17}: the 17.10~\AA \ radiative decay (downward-pointing
wavy line), collisional excitations (upward-pointing lines), and
photoexcitations (upward-pointing wavy lines). Since the $2p^53s\>
(J=2)$ level is the first excited level in \ion{Fe}{17}, we denote
it by the subscript ``2.'' The numbers associated with the lines are
respectively the radiative decay rate $A_{21}$, the collision rate
coefficients $\gamma _{2j}$ for $T_{\rm e}=4$ MK, and the oscillator
strengths $f_{2j}$ (we report $\sum_{j=6}^{15}\gamma _{2j}$ and
$\bar f =\sum_{j=6}^{15}\lambda _{2j} f_{2j}/\sum_{j=6}^{15}\lambda 
_{2j}$ for the ten transitions into the $2p^53p$ manifold). Because of
the large effective oscillator strength for photoexcitations into the
$2p^53p$ manifold, the population of the $2p^53s\> (J=2)$ level, and
hence the strength of the 17.10~\AA \ emission line, can be significantly
reduced if the photoexcitation continuum is sufficiently strong in the
190--410~\AA \ waveband responsible for the $2p^53s\> (J=2)$--$2p^53p$
transitions. In He-like ions, it is the strength of the continuum at the
wavelengths of the $1s2s\> ^3S_1$--$1s2p\> ^3P_{2,1}$ transitions which
affects the $f/i$ line ratio. For O, Ne, Mg, Si, S, Ar, Ca, and Fe 
these wavelengths are 1623, 1263, 1036, 865, 743, 637, 551, and 404~\AA ,
respectively. Because the relevant waveband for \ion{Fe}{17} is at
shorter wavelengths/higher energies than those of He-like ions, the
\ion{Fe}{17} $I(17.10~{\rm \AA})/I(17.05~{\rm \AA})$ line ratio is
considerably less sensitive to photoexcitation.

We used LXSS to calculate \ion{Fe}{17} spectral models for a blackbody
photoexcitation continuum, densities $n_{\rm e}= 10^{10}$--$10^{16}~\rm
cm^{-3}$, and temperatures $T_{\rm e} =2$--8~MK (spanning the range for
which the \ion{Fe}{17} ionization fraction is $\gax 0.1$;
\citealt{maz98}). Assuming a white dwarf mass $\Mwd = 0.49\, \Msun $
($\Rwd=9.8\times 10^8$~cm) and distance $d= 100$~pc, \citet{mau99} found
that the FUV flux density of EX~Hya could be explained if the entire
white dwarf surface radiates with a blackbody spectrum with a temperature
$T_{\rm bb}\approx 27$~kK. Consequently, we assumed for our first set
of models that $T_{\rm bb}=30$~kK and $T_{\rm e}=2$, 4, and 8~MK. The
results of these calculations are shown in the left panel of Figure~3.
It is seen that the $R(15.01~{\rm \AA })$ line ratio shows the most
temperature sensitivity, while the $R(15.45~{\rm \AA })$ line ratio shows
the least. The theoretical ratios are compared to the observed ratios for
EX~Hya and the Sun in the middle panel of the Figure~3. As with other
theoretical models of \ion{Fe}{17} \citep{bro01}, we are unable to
reproduce the observed $R(15.01~{\rm \AA })$ [hence the $I(15.01~{\rm 
\AA })/ I(15.26~{\rm \AA })$] line ratio, although the value we measure
for this ratio is consistent with that measured in the Sun. The observed
$R(15.26~{\rm \AA })$ and $R(15.45~{\rm \AA })$ line ratios are as
predicted by our models, but the $R(17.05~{\rm \AA })$ and $R(17.10~{\rm
\AA })$ line ratios are too high and too low, respectively, unless the
plasma density $n_{\rm e}\gax 3\times 10^{14}~\rm cm^{-3}$.

It is alternatively possible that the observed $R(17.10~{\rm \AA })$
line ratio could be explained if the effective temperature of the
photoexcitation continuum is greater than assumed for the set of models
discussed above. Indeed, following \citet{gan95}, \citet{mau99} proposed
that the spin-phase modulation of the FUV flux of EX~Hya is the result
of the varying aspect of a $T_{\rm bb}\approx 37$~kK hotspot on a
$T_{\rm bb}\approx 20$~kK white dwarf. To investigate the consequences
of a hotter photoexcitation continuum, we assumed for our second set of
models that $T_{\rm e}=4$~MK (the peak of the \ion{Fe}{17} ionization
fraction) and $T_{\rm bb}=30$, 40, 50, and 60~kK. The results of these
calculations are shown in the right panel of Figure~3. It is seen that
the $R(17.05~{\rm \AA })$ and $R(17.10~{\rm \AA })$ line ratios are
sensitive to the temperature of the photoexcitation continuum, whereas
the $R(15.26~{\rm \AA })$ and $R(15.45~{\rm \AA })$ line ratios are not.
The $R(17.10~{\rm \AA })$ line ratio is driven toward zero (the
``high-density limit'') for blackbody temperatures $T_{\rm bb}\gax 60$~kK.

To demonstrate the level of sensitivity to photoexcitation of the
$I(17.10~{\rm \AA })$ line, we show in Figure~4 a plot of the
$I(17.10~{\rm \AA })/I(17.05~{\rm \AA })$ line ratio as a function of
density assuming a plasma temperature $T_{\rm e}=4$~MK and blackbody
temperatures $T_{\rm bb}=20$, 30, 35, 40, 45, 50, 55, and 60~kK. The
curve for $T_{\rm bb}=20$ kK is virtually identical to the curve without
photoexcitation. As the blackbody temperature increases, the line ratio
at low densities is seen first to increase slightly and then decrease
strongly, approaching the high-density limit for $T_{\rm bb} \gax 60$~kK.
If the low $I(17.10~{\rm \AA })/I(17.05~{\rm \AA })$ line ratio observed
in EX~Hya is due solely to photoexcitation, the blackbody temperature
$T_{\rm bb}\gax 55$~kK. With $N_{\rm H}\approx 10^{20}~\rm cm^{-2}$ there
is no observable flux in the EUV for any of these models, so the only
constraint is that they not produce too much flux in the UV. Assuming
that the maximum 1010~\AA \ flux density is $2.5\times 10^{-13}~\rm
erg~cm^{-2}~s^{-1}~\AA ^{-1}$ \citep{mau99}, the fractional emitting
area of these hotspots must be $f\le 5.8\%$, 2.6\%, 2.0\%, and 1.5\%
for $T_{\rm bb}= 40$, 50, 55, and 60~kK, respectively.

\section{Summary and Discussion}

We have shown that the \ion{Fe}{17} $I(17.10~{\rm \AA })/I(17.05~{\rm
\AA })$ line ratio observed in the {\it Chandra\/} HETG spectrum of 
EX~Hya is significantly smaller than that observed in the Sun or other
late-type stars. Using LXSS, a new plasma code based on HULLAC atomic
data, we find that the observed line ratio can be explained if the
plasma density $n_{\rm e}\gax 3\times 10^{14}~\rm cm^{-3}$. However,
if photoexcitation is included in the level population kinetics, the
line ratio can be explained for any density if the photoexcitation
temperature $T_{\rm bb}\gax 55$~kK. This latter model is consistent with
the assumptions (blackbody emitter, dilution factor equal to one-half)
and the observed UV flux density only if this hotspot, and the overlying
volume of million-degree plasma, covers a very small fraction of the
surface area of the white dwarf: $f\lax 0.02$. Such a hotspot is smaller
and hotter than is usually inferred from optical and UV light curves, 
but it is possible that the accretion-heated photosphere of the white
dwarf contains a range of temperatures, with the required $T_{\rm bb}
\gax 55$~kK spot applying only to that unfortunate piece of real estate
lying directly beneath the accretion column. A surrounding larger ($f
\approx 0.1$) and cooler ($T_{\rm bb}\approx 37$~kK) suburb might then
produce the flux modulations observed at longer wavelengths, but to
account for {\it its\/} contribution to the UV flux density, the hottest
part of the spot would have to be even smaller. The mean 0.5--10 keV
flux of EX~Hya during our observation was $f_{\rm X}\approx 1\times
10^{-10}~\rm erg~cm^{-2}~s^{-1}$, and with a white dwarf mass $\Mwd =
0.49\, \Msun $ ($\Rwd=9.8\times 10^8$~cm) and distance $d=100$~pc,
this implies an X-ray luminosity $L_{\rm X}\approx 1\times 10^{32}~\rm
erg~s^{-1}$, hence a mass-accretion rate $\Mdot= 2\Rwd L_{\rm X}/G\Mwd
\approx 4\times 10^{15}~\rm g~s^{-1}$, hence a post-shock density 
$n\ge \Mdot /4\pi f\Rwd ^2\mu m_{\rm H} (v_{\rm ff} /4)\approx 2\times
10^{14}~\rm cm^{-3}$ for $f\lax 0.02$. Either way, then, the {\it
Chandra\/} HETG spectrum of EX~Hya requires a plasma density which is
orders of magnitude greater than that observed in the Sun or other
late-type stars. The \ion{Fe}{17} $I(17.10~{\rm \AA })/I(17.05~{\rm \AA
})$ density diagnostic is useful in sources in which the efficacy of the
He-like density diagnostics is compromised by the presence of a bright UV
continuum.

\acknowledgments

We thank H.~Tananbaum for the generous grant of Director's Discretionary
Time which made these observations possible. We gratefully acknowledge
J.~Raymond and N.~Brickhouse for ongoing discussions about high-density
plasma in EX~Hya, P.~Beiersdorfer and G.~Brown for discussions about 
\ion{Fe}{17}, J.~Saba for providing the solar \ion{Fe}{17} line ratios
listed in Table~1 and shown in Fig.~3, and the referee for a number of
suggestions which improved the clarity of the manuscript. C.~M.\ and 
D.~L.\ were supported in part by NASA Long-Term Space Astrophysics
Program grant S-92654-F and NASA {\it Chandra\/} Guest Observer grant
NAS8-39073. This work was performed under the auspices of the U.S.\
Department of Energy by University of California Lawrence Livermore
National Laboratory under contract No. W-7405-Eng-48.

\clearpage 


\clearpage 


\begin{deluxetable}{lcc}
\small
\tablecolumns{3} 
\tablewidth{0pc} 
\tablenum{1}
\tablecaption{\ion{Fe}{17} Line Ratios}
\tablehead{
\colhead{Ratio} &
\colhead{EX~Hya} &
\colhead{Sun}}
\startdata
\hbox to 1.8in{$I(15.01~{\rm \AA })/I(16.78~{\rm \AA })$\leaders\hbox to 0.5em{\hss.\hss}\hfill}& $1.23\pm 0.15$& $1.04\pm 0.13$\\
\hbox to 1.8in{$I(15.26~{\rm \AA })/I(16.78~{\rm \AA })$\leaders\hbox to 0.5em{\hss.\hss}\hfill}& $0.50\pm 0.09$& $0.51\pm 0.08$\\
\hbox to 1.8in{$I(15.45~{\rm \AA })/I(16.78~{\rm \AA })$\leaders\hbox to 0.5em{\hss.\hss}\hfill}& $0.05\pm 0.06$& $\cdots      $\\
\hbox to 1.8in{$I(17.05~{\rm \AA })/I(16.78~{\rm \AA })$\leaders\hbox to 0.5em{\hss.\hss}\hfill}& $1.65\pm 0.20$& $1.40\pm 0.20$\\
\hbox to 1.8in{$I(17.10~{\rm \AA })/I(16.78~{\rm \AA })$\leaders\hbox to 0.5em{\hss.\hss}\hfill}& $0.08\pm 0.07$& $1.32\pm 0.14$\\
\hbox to 1.8in{$I(15.26~{\rm \AA })/I(15.01~{\rm \AA })$\leaders\hbox to 0.5em{\hss.\hss}\hfill}& $2.46\pm 0.42$& $2.02\pm 0.28$\\
\hbox to 1.8in{$I(17.10~{\rm \AA })/I(17.05~{\rm \AA })$\leaders\hbox to 0.5em{\hss.\hss}\hfill}& $0.05\pm 0.04$& $0.93\pm 0.11$\\
\enddata
\end{deluxetable}

\clearpage 


\begin{figure}
\figurenum{1}
\epsscale{1.0}
\plotone{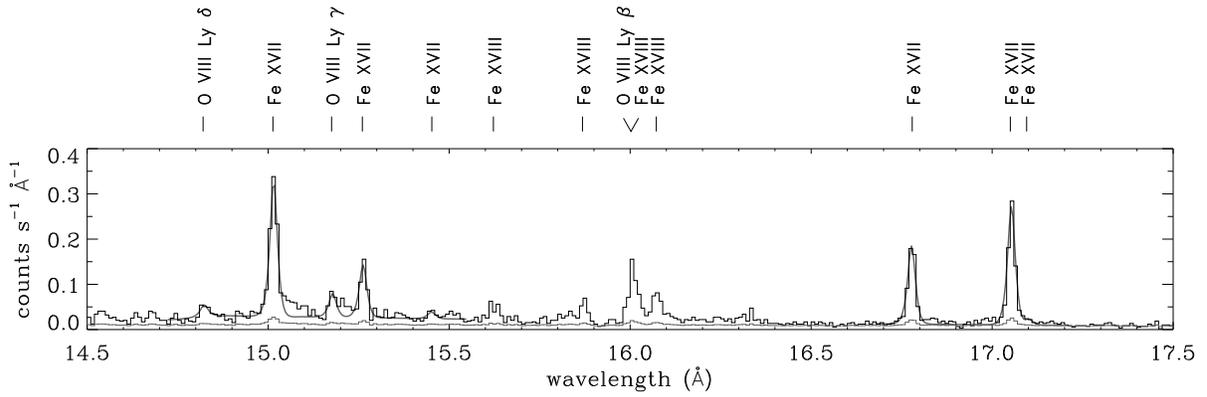}
\caption{
Detail of the {\it Chandra\/} MEG spectrum of EX~Hya. Data are shown
by the upper dark histogram, the $1\, \sigma $ error vector by the lower 
gray histogram, and the model fit ($\lambda\lambda = 14.72$--15.55~\AA \
and 16.68--17.20~\AA ) by the thick gray curve. Strongest emission
lines of \ion{Fe}{17}, \ion{Fe}{18}, and \ion{O}{8} are labeled along
the top of the figure. Data combines $\pm 1$st orders and is binned to
0.01~\AA .}
\end{figure}
\clearpage 

\begin{figure}
\figurenum{2}
\epsscale{0.537}
\plotone{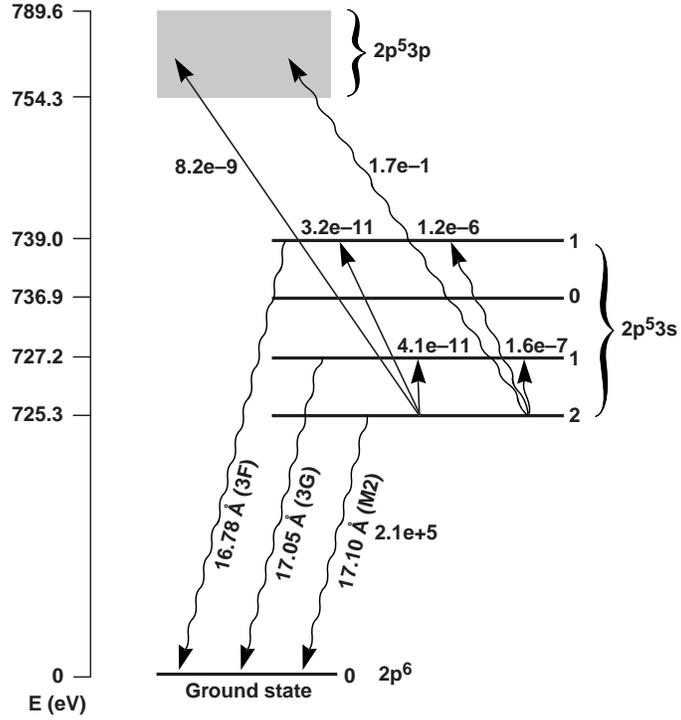}
\caption{
Level population processes in Ne-like \ion{Fe}{17}. We show the
$2p^6$--$2p^53s$ radiative decays (downward-pointing wavy lines) and
the dominant collisional excitation (upward-pointing lines) and
photoexcitation (upward-pointing wavy lines) paths out of the $2p^53s\>
(J=2)$ level. Numbers associated with the lines are respectively the
radiative decay rate $A_{21}$, collision rate coefficients $\gamma _{2j}$
for $T_{\rm e}=4$ MK, and oscillator strengths $f_{2j}$.}
\end{figure}
\clearpage 

\begin{figure}
\figurenum{3}
\epsscale{1.0}
\plotone{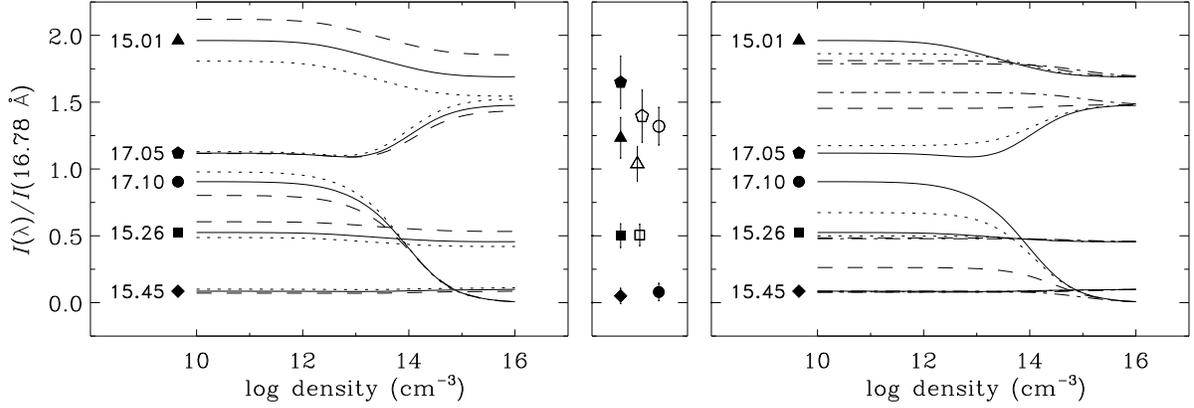}
\caption{
{\it Left panel\/}: LXSS model \ion{Fe}{17} line ratios as a function
of density for a 30 kK blackbody photoexcitation continuum and plasma
temperatures $T_{\rm e}=2$, 4, and 8 MK (dotted, continuous, and dashed
curves, respectively). {\it Middle panel\/}: Line ratios measured in
EX~Hya (filled symbols) and the Sun (open symbols). {\it Right panel\/}:
LXSS model \ion{Fe}{17} line ratios as a function of density for a plasma
temperature $T_{\rm e} =4$ MK and a blackbody photoexcitation continuum
with $T_{\rm bb}=30$, 40, 50, and 60~kK  (continuous, dotted, dashed, and
dot-dashed curves, respectively).}
\end{figure}

\begin{figure}
\figurenum{4}
\epsscale{0.537}
\plotone{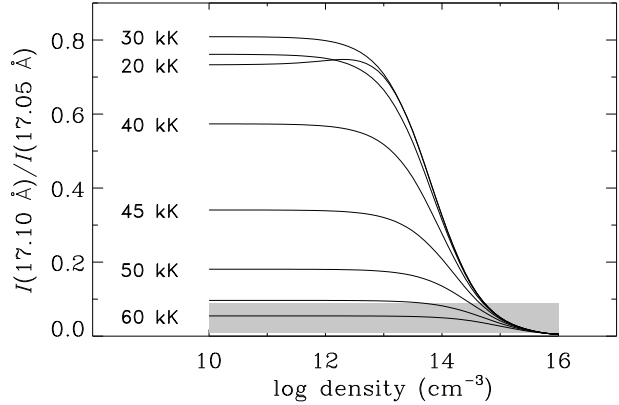}
\caption{
LXSS model \ion{Fe}{17} $I(17.10~{\rm \AA })/I(17.05~{\rm \AA })$ line
ratio as a function of density for a plasma temperature $T_{\rm e}=4$ MK
and a blackbody photoexcitation continuum with $T_{\rm bb}=20$, 30, 35,
40, 45, 50, 55, and 60~kK. Grey stripe delineates the $1\, \sigma $
envelope of the ratio measured in EX~Hya.}
\end{figure}
\clearpage 

\end{document}